\documentclass{WileyMSP-template}

\usepackage{graphicx}%
\usepackage{multirow}%
\usepackage{amsmath,amssymb,amsfonts}%
\usepackage{amsthm}%
\usepackage{mathrsfs}%
\usepackage[title]{appendix}%
\usepackage[dvipsnames]{xcolor}
\usepackage{textcomp}%
\usepackage{manyfoot}%
\usepackage{booktabs}%
\usepackage{algorithm}%
\usepackage{algorithmicx}%
\usepackage{algpseudocode}%
\usepackage{listings}%
\usepackage{makecell}
\usepackage{bm}
\usepackage{threeparttable}

\usepackage{mathtools}
\usepackage{esint}

\usepackage{float}

\usepackage{hyperref}
\hypersetup{
    colorlinks,
    citecolor=blue,
    filecolor=blue,
    linkcolor=blue,
    urlcolor=blue
}
 
\newcommand*{\blue}{\textcolor{blue}}

\newcommand{\beg}{\begin{equation}}
\newcommand{\en}{\end{equation}}

 \newcommand{\lam}{\lambda}

\newcommand{\eref}[1]{Equation~(\ref{#1})}
\newcommand{\re}[1]{(\ref{#1})}

\newcommand{\esref}[1]{Equation.~(\ref{#1})}

\newcommand{\be}{\begin{eqnarray}}
\newcommand{\ee}{\end{eqnarray}}

\newcommand{\bs}{\begin{equation}\begin{split}}
\newcommand{\es}{\end{split}\end{equation}}

\begin{document}

\pagestyle{fancy}
\rhead{\includegraphics[width=2.5cm]{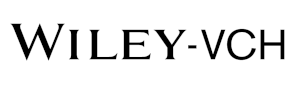}}

\title{Fundamental limits on the electron-phonon coupling and superconducting $T_c$  }

\maketitle


\author{Dmitrii V.  Semenok*}
\author{Boris L. Altshuler}
\author{Emil A. Yuzbashyan*}

 \begin{affiliations}
Dmitrii V.  Semenok\\
 Center for High Pressure Science \& Technology Advanced Research (HPSTAR),  Bldg. 8E, ZPark, 10 Xibeiwang East Rd,  Beijing,  Haidian  100193,  China\\
dmitrii.semenok@hpstar.ac.cn\\

Boris L. Altshuler\\
 Physics Department, Columbia University,  538 West 120th Street,  New York,  NY  10027,  USA\\
 
 Emil A. Yuzbashyan\\
 Department of Physics and Astronomy, Center for Materials Theory,  Rutgers University,   Piscataway, NJ  08854,  USA\\
 eyuzbash@physics.rutgers.edu

\end{affiliations}



\begin{abstract}

\begin{abstract}
 \begin{abstract}
{Fundamental upper bounds on the electron-phonon interaction strength and superconducting transition temperature \( T_c \) in metals are established based on the intrinsic instability of the equilibrium between electrons and the crystal lattice under strong interaction. This instability   explains why observed electron-phonon coupling constants are limited to \( \lambda \lesssim 4 \). The theory also accounts for the mechanism of metastable superconductivity with enhanced \( T_c \), which emerges near the instability threshold. Based on theoretical analysis and comparison with experimental data, room-temperature phonon-mediated superconductivity is found to be feasible exclusively in hydrogen compounds.}
\end{abstract}

\end{abstract}

\end{abstract}


 \section{Introduction}\label{sec1}

  The interaction of electrons with lattice vibrations (phonons) is at the heart of the theory of metals.  It  determines  superconducting  $T_c$,  electrical and thermal conductivities,     and many other  physical properties~\cite{abrikosov,grimvall}. Existing theoretical frameworks, such as the Bardeen-Cooper-Schiffer (BCS)~\cite{bcs}, Migdal~\cite{migdal}, and Eliashberg~\cite{eli1st} theories, provide no limits on the electron-phonon interaction   and, as a result,   on $T_c$. In this paper, we show that there are  in fact intrinsic upper bounds on  both these quantities dictated by the stability of the metal with respect to the electron-lattice interaction. We compare our results with experimental data and argue that   room-temperature phonon-mediated superconductivity is entirely realistic but only in  hydrogen-rich compounds.

\begin{figure}[t!]
\centerline{\includegraphics[width=0.92\columnwidth]{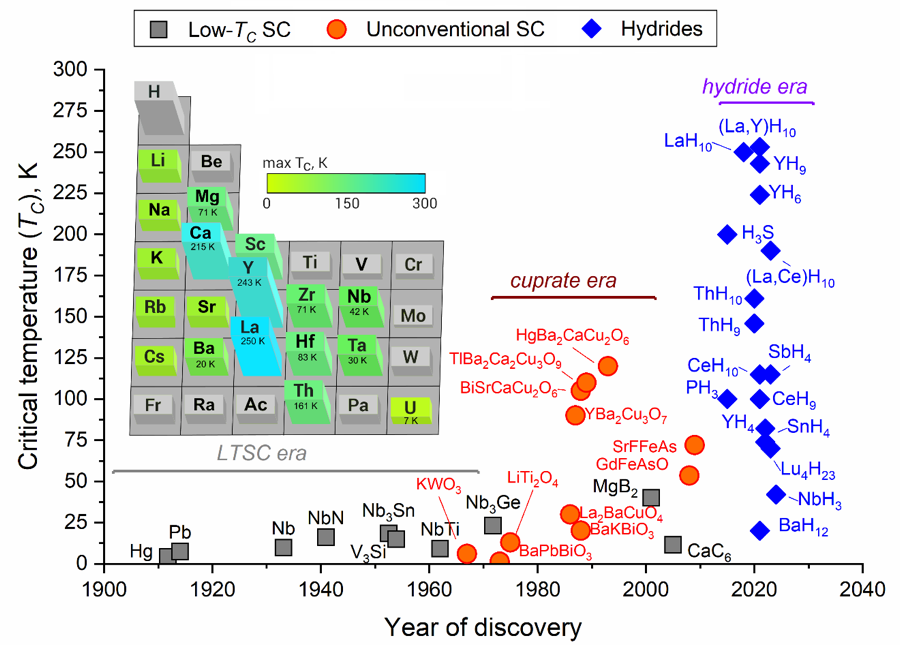}}
\caption{The evolution of  superconducting materials in 20th and 21st centuries as exemplified by metals and intermetallic alloys (grey squares), which are well described by the BCS theory, unconventional superconductors (orange circles) and hydrides (blue rhombi). The inset shows the distribution of superconducting transition temperatures for the hydrides of various metals.  Shown in grey are the metals whose polyhydrides  have not been investigated for superconductivity.    }
\label{fig0} 
\end{figure}

  The question, ``What is the maximum possible superconducting $T_c$?" has remained unanswered since the discovery of  superconductivity by  Onnes in 1911~\cite{onnes}, despite remarkable progress in this area~\cite{anderson,leavens,mitrovic, kivelson1, kivelson2,jiang,omega_max,sadovskii}. At the same time,  the $T_c$ of real materials never exceeded 133 K at atmospheric pressure and 160 K at elevated pressures ($\sim$30 GPa) in   more than a hundred years (1911-2011) of experimental experience. 
It is believed that  metallic hydrogen  is a superconductor with one of  the highest  critical temperatures~\cite{wigner,ashcroft}. This is because $T_c$  is proportional to  lattice vibration frequencies, which are highest in this material since hydrogen is the lightest element. Unfortunately, producing  metallic hydrogen requires pressures in excess of 450 GPa~\cite{eremets,loubeyre},  at the limit of reach of current experimental techniques for transport measurements. There is, however, an ingenious solution---alloying  hydrogen with other elements~\cite{zurek}. This provides an effective chemical pressure thus reducing the external pressure necessary to produce a stable metal.

And indeed, compressed  polyhydrides became the leaders in the quest for the highest $T_c$ since the discovery of record superconductivity in 2014 and 2018  in the cubic hydrides of sulfur (H$_3$S, max $T_c=200$~K) and lanthanum (LaH$_{10}$, max $T_c=250$~K). This ``hydride revolution'' gave rise to justified hopes   of room-temperature superconductivity in the near future, see \textbf{Figure~\ref{fig0}}. Moreover, several incorrect reports of  $T_c$ exceeding room temperature  in the ternary systems C-S-H~\cite{snider1,snider2}, Y-Pd-H~\cite{snider2}, and Lu-N-H~\cite{gammon}   emerged but were quickly refuted. At the present moment, the maximum well reproduced critical temperature is $T_c = 250$~K in LaH$_{10}$, (La,Y)H$_{10}$~\cite{drozdov5,Somayazulu,sun}, (La, Sc)H$_{10}$~\cite{dima}, and other ternary compounds containing lanthanum. At the same time, the question whether room-temperature superconductivity is attainable remains open.

Here we will approach this problem from a different perspective. Our recent studies within the Migdal-Eliashberg (ME) theory revealed  that the metallic and superconducting states are unstable with respect to strong electron-phonon interaction~\cite{emil,kinetic}. A traditional measure of the strength of this interaction is the electron-phonon coupling constant $\lam$  defined as the sum of the couplings $\lam_k$   to the individual lattice vibration modes. This   characterization of the  interaction strength  with a single number is   not unique, and   we introduce below another metric $\xi$---the \textsl{stability parameter} of the metal.  It corresponds to the contribution of the electron-phonon interaction to the electronic specific heat and is also a sum of $\lam_k$ but with unequal weights.  

Indeed, we will see that the dynamical stability of a metal requires $\xi <\xi_*=1$. The metallic state ceases to be the global minimum of the free energy at a smaller value  $\xi_c<\xi_*$ via a first order phase transition and is no longer even a local minimum when $\xi>\xi_*$. For a fixed phonon spectrum, $\lam$ and $\xi$ are directly proportional to each other, so that this condition translates into $\lam < \lam_c<\lam_*$. The value of $\lam_*$ is smallest, $\lam_*= 3.69$, for dispersionless (Einstein)  phonons. When $\lam$ exceeds $\lam_c$ ($\xi$ exceeds $\xi_c<1$), the crystal structure of the metal becomes unstable and is either destroyed or undergoes a reconstruction  lowering the value of $\lam$ below $\lam_c$. This result provides an effective tool to investigate the maximum    transition temperature of electron-phonon superconductors thanks to an upper bound on $T_c$ within the ME theory in terms of $\lam$ and the average  square phonon frequency~\cite{mitrovic,allen}.

While our study explicitly addresses phonon-mediated superconductivity, the stability constraints we identified hinge fundamentally on the retarded nature of boson-mediated interactions. Retardation, quantified by the ratio of the characteristic bosonic frequency to the Fermi energy   is crucial. Specifically, our theoretical results   are most robust when this ratio is small, indicating substantial retardation. Our arguments remain valid even  if the mediating bosonic field is effectively an order parameter constructed from fermions themselves. With this consideration, the conceptual approach presented should, in principle, be adaptable to   cuprates and other unconventional superconductors, provided a well-defined model of boson-mediated pairing exists for these systems. However,  significant modifications  to our theoretical framework are necessary when characteristic bosonic frequencies become comparable to the fermionic bandwidth.

\section{Analysis of experimental data}\label{sec2}

Presently, superconductors with highest critical temperatures are the polyhydrides. They also have some of the highest values of the electron-phonon interaction constant $\lam\sim 2-3$ and of the stability parameter $\xi\sim 0.2-0.5$ (\textbf{Table~\ref{sc_table}}). Interestingly, the values of $\lam$   for hydrides lie approximately within the same range as for soft superconducting metals Pb and Bi and their alloys, even though the critical temperatures   are tens of times larger. This begs the question: Can the electron-phonon interaction be arbitrarily large  or is it inherently bounded from above?

\begin{figure}[h!]
\centerline{\includegraphics[width=0.7\columnwidth]{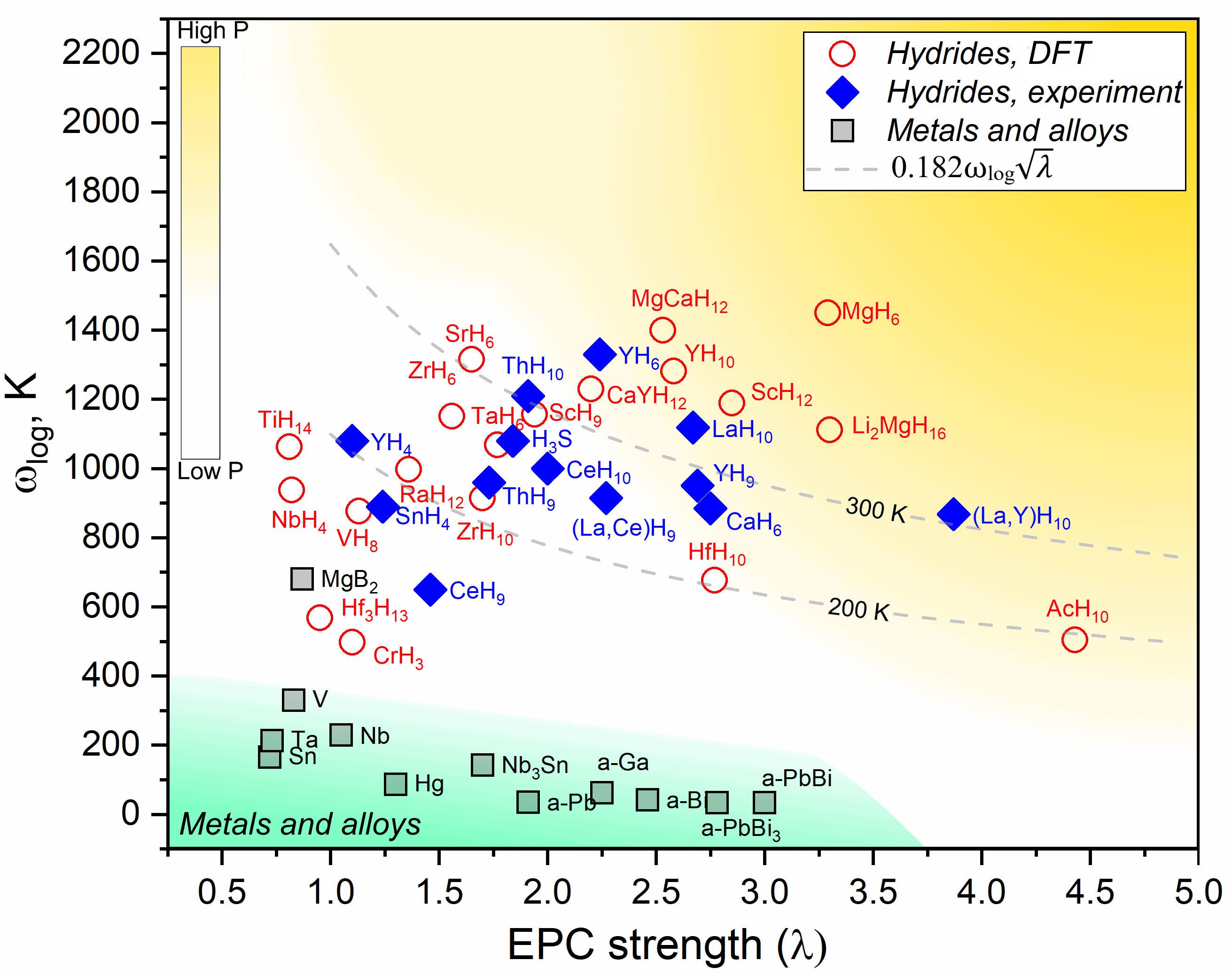}}
\caption{Distribution of superconducting materials in the $\lam - \omega_\mathrm{log}$ plane. (a) Open red circles and filled blue rhombi show  DFT (Density Functional Theory) predicted and experimentally synthesized hydrides, respectively. Filled grey squares correspond to simple metals, alloys, and MgB$_2$. Dashed grey lines are defined by the equation $T_c^{\mathrm{emp}}\equiv0.182\omega_\mathrm{log}\sqrt{\lam}=\mathrm{const}$, where $T_c^{\mathrm{emp}}$ is the empirical strong coupling estimate for $T_c$~\cite{allen}.   }  \label{fig1} 
\end{figure}

\begin{figure}[h!]
\centerline{\includegraphics[width=0.92\columnwidth]{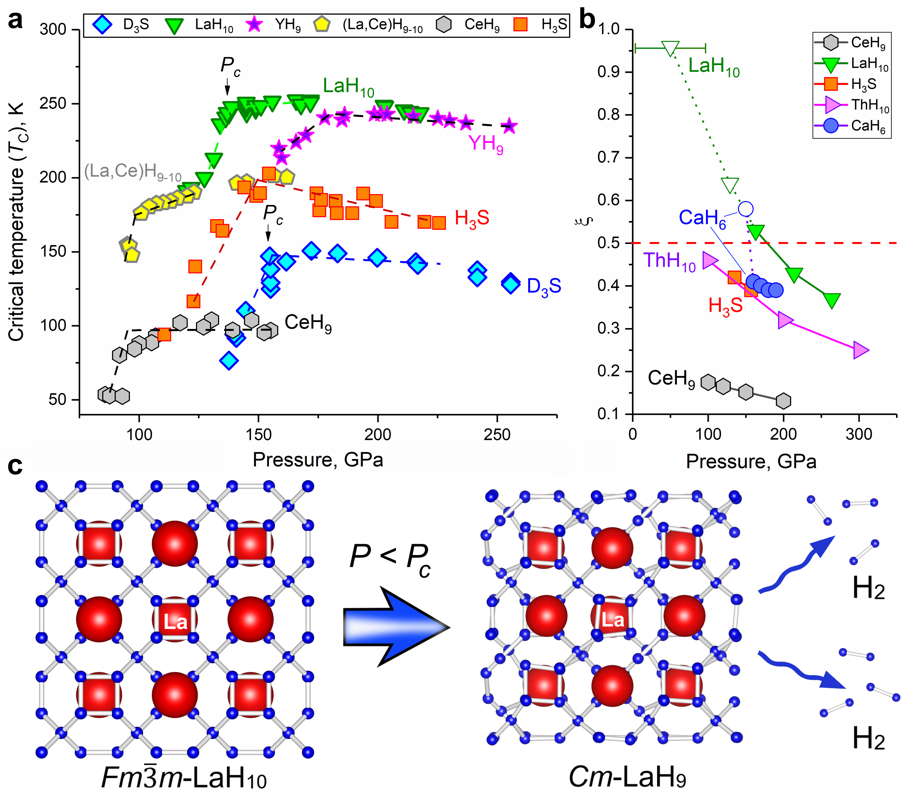}}
\caption{Instability and decomposition of polyhydrides with decreasing pressure. (a) Experimentally observed dependence of the critical temperature $T_c$ on pressure for certain polyhydrides and deuterium sulfide (D$_3$S). As the pressure is lowered below its critical value $P_c$, $T_c$ abruptly decreases. This is accompanied by a distortion of the crystal structure and partial loss of hydrogen. (b) Dependence of the stability parameter $\xi$ on pressure  for several experimentally synthesized hydrides.   The ${\xi=0.5}$ line marks  the supposed first order phase transition, i.e., upon crossing this line, the destruction of the structure is merely a function of time and the height of the kinetic barrier.  Filled symbols correspond to experimental data, empty ones are DFT calculations.  The point ${\xi= 0.96}$ (LaH$_{10}$) is an extrapolation obtained by  multiplying the Eliashberg function $\alpha^2F(\omega)$ for LaH$_{10}$ at 129 GPa by a factor of 1.5. (c) Decomposition of  LaH$_{10}$  upon lowering the pressure below ${P_c = 138}$ GPa  accompanied by a distortion of its cubic structure,  partial loss of hydrogen (H$_2$) and formation of lower hydrides, such as LaH$_9$.  }
\label{fig2} 
\end{figure}

 Hydride  superconductors or superhydrides  (Figs.~\ref{fig1} and \ref{fig2}) can be synthesized only at high pressure $P$, about ${100-200}$ GPa, in special diamond anvil cells. Subsequent lowering of the pressure   leads to a domelike dependence of $T_c$ on   $P$ characteristic of hydrides~\cite{drozdov5,Somayazulu,kong,chen,drozdov1}.  Increasing $P$ does not  lead to the destruction of the polyhydrides but is accompanied by an increase of  phonon frequencies, decrease of the electron-phonon interaction strength, and, most importantly, decrease of $T_c$. An appropriate measure of the magnitude of phonon frequencies at these $\lam$~\cite{allen} is the average logarithmic frequency $\omega_\mathrm{log}$ defined as $\ln \omega_\mathrm{log}=\frac{1}{\lam}\sum_k \lam_k \ln\omega_k$.

 The situation seen when decreasing the pressure is almost the opposite---phonons soften, while $\lam$ and $\xi$ grow (\textbf{Figure~\ref{fig2}\blue{b}}). It may seem that decreasing the pressure further and thus increasing $\lam$, we can make $T_c$ arbitrarily large, since   $T_c \propto \sqrt{\lam}$ at large $\lam$  according to the ME theory~\cite{allen,combescot}.  
 In reality,  this turns out to be impossible, since the growth of the electron-phonon interaction   in hydrides inevitably leads to the distortion of the crystal structure, lowering of its symmetry,  diffusion and partial loss of hydrogen, and, as a result, to an abrupt lowering of the superconducting  $T_c$ (Figure~\ref{fig2}\blue{a}). Good examples of this process are the decompression of D$_3$S~\cite{drozdov1,minkov1}, LaH$_{10}$~\cite{drozdov5}, YH$_9$~\cite{kong}, and CeH$_9$~\cite{chen} as well as recent studies of ternary lanthanum-cerium superhydrides  (La,Ce)H$_9$~\cite{chen3,bi1}. 
 
 \section{Fundamental limit on the electron-phonon interaction strength   in metals}
 
 When does a metal stop being a metal as the pull between conduction electrons and phonons increases?  In a new study~\cite{emil}, we discovered that the specific heat   of  conduction electrons ($C_\mathrm{el}$) turns negative when the electron-phonon coupling constant exceeds a certain threshold $\lam_*$. The precise value of $\lam_*$ depends on the phonon spectrum and, in particular, $\lam_*=3.69$ for Einstein phonons. Stability analysis of the kinetic equations  (Section~\ref{sec:stability_analysis}, see also Sections~\ref{sec:proof}, \ref{sec:polaronic},
 and Ref.~\cite{kinetic}) shows that when
 $C_\mathrm{el}<0$, the metal is  unstable with respect to an infinitesimal difference between the electron and phonon temperatures, $T_\mathrm{el}$ and $T_\mathrm{ph}$, respectively. Specifically, if initially $T_\mathrm{el}>T_\mathrm{ph}$,
 $T_\mathrm{el}$ will grow exponentially and the system will never equilibrate.  This indicates that    metals with $\lam>\lam_*$ cannot exist in nature even in a metastable state. 
    
  This result does not rely on effective electron-phonon models, such as Holstein or Fr\"ohlich Hamiltonians. Such simplified models inherently exhibit artificial phonon softening at a \textsl{bare} electron-phonon coupling $\lambda_0 \approx 0.5$~\cite{migdal,eli1st,agd}, resulting from double counting of the static electronic contribution (overscreening)~\cite{kagan,geilikman,screen}. Most importantly, this spurious lattice instability does not define a meaningful upper limit on the \textsl{physical} electron-phonon coupling $\lambda$, since $\lambda \to \infty$ as $\lambda_0 \to 0.5$. In contrast, our approach provides a rigorous upper bound valid for any phonon spectrum without requiring phonon softening (see also Section~\ref{sec:polaronic}). In particular, this upper bound cannot be circumvented by invoking multiple phonon modes with comparable coupling strengths~\cite{kivelson2}, as this does not alter the ME expression for $C_\mathrm{el}$.

 As  noted above, the stability requirement $C_\mathrm{el}\ge 0$ is equivalent to a certain upper limit on the electron-phonon interaction. To see this, consider a well-known expression for the electronic specific heat at temperature $T$~\cite{prange,lee,dolgov,golubov_1} (in units $\hbar=k_B=1$),
 \beg
C_\mathrm{el}=  \frac{2}{3}\pi^2 \nu_0 T\left[ 1-  \int_0^\infty g\left( \frac{\omega}{2\pi T}\right)\frac{2  \alpha^2 F(\omega)}{\omega} d\omega\right]\! ,
\label{C}
\en
 where $\nu_0$ is the density of states at the Fermi energy,  
$g(x)=6 x^2+12x^3\mathrm{Im} \psi'(\i x)+6x^4\mathrm{Re} \psi'' (\i x),$ and
$\psi(x)$ is the digamma function.
The combination $\alpha^2 F(\omega)$ is  the Eliashberg function defined as
\beg
\alpha^2 F(\omega)= \sum_{k} \frac{\lam_k \omega_k}{2} \delta(\omega-\omega_k),
\en
where  $\omega_k$ are the  frequencies of lattice vibration modes.

With the help of Equation~\eref{C}, the condition for stability, \( C_\mathrm{el} > 0 \) for all \( T > T_c \), takes the form
\begin{equation}
\xi \equiv \max_T \left\{ \int_0^\infty g\left( \frac{\omega}{2\pi T} \right) \frac{2 \alpha^2 F(\omega)}{\omega} \, d\omega \right\} < 1.
\label{C>0}
\end{equation}
Here, the maximum is taken over temperature \( T \). The \textsl{stability parameter} \( \xi \) is another measure of the electron-phonon interaction strength. Like \( \lambda \), it scales linearly with the overall magnitude of the Eliashberg function, but it is instead a weighted average of the mode-resolved couplings \( \lambda_k \),
\begin{equation}
\xi = \sum_{k=1}^N \lambda_k G_k, \qquad G_k = g\left( \frac{\omega_k}{2\pi T_{\max}} \right),
\end{equation}
where \( T_{\max} \) is the temperature at which the integral in Equation~\re{C>0} reaches its maximum.

 \begin{figure}[t!]
\centerline{\includegraphics[width=0.92\columnwidth]{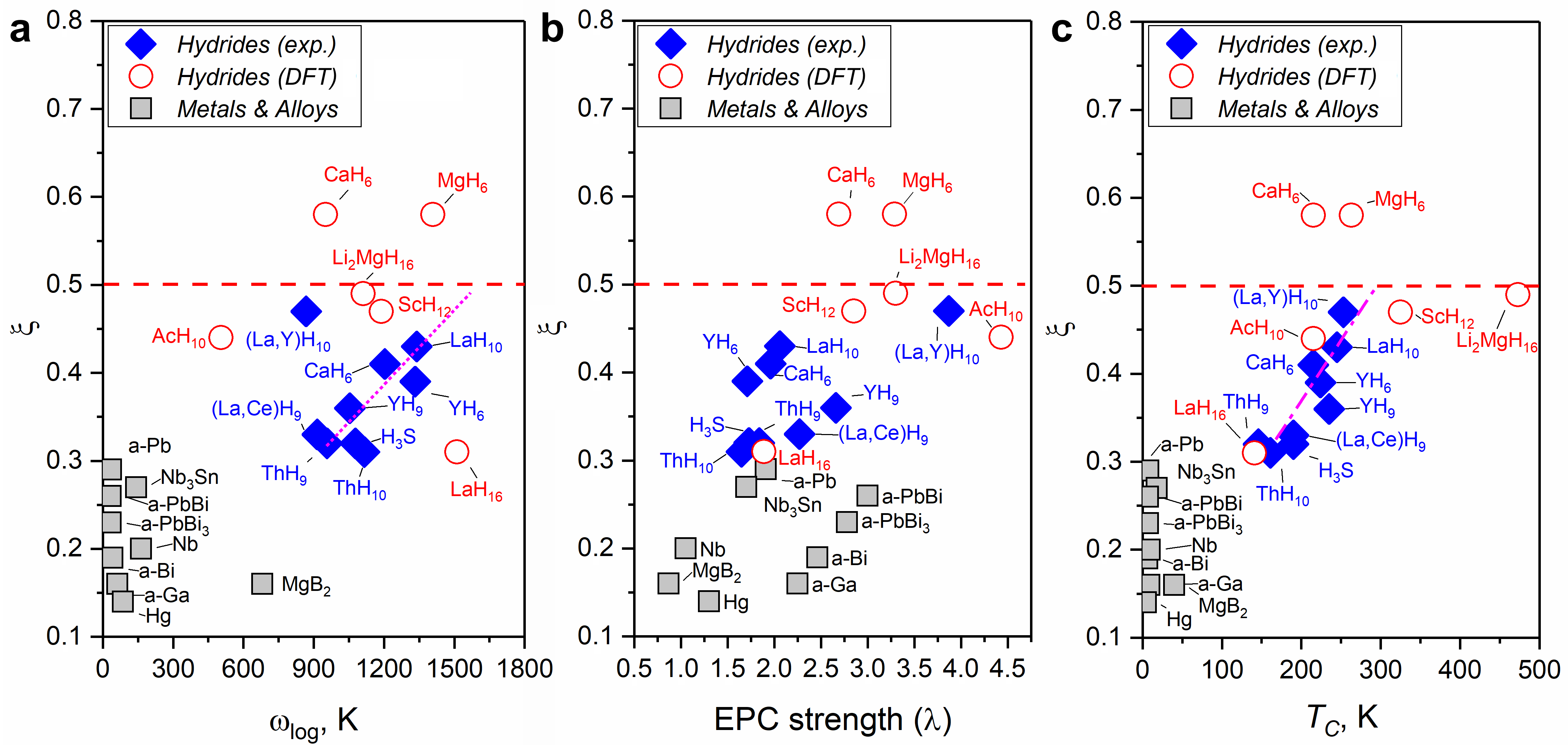}}
\caption{Values of the stability parameter $\xi$ for various hydrides as a function of (a) average logarithmic  frequency $\omega_\mathrm{log}$, (b) electron-phonon coupling constant $\lam$, and (c) superconducting $T_c$. Filled and empty symbols correspond to experiment and DFT calculations, respectively.  Dashed violet lines indicate the trend  apparent  among the best hydride superconductors.}
\label{figS1} 
\end{figure}

 The   phase transition from the metal (superconductor below $T_c$) to a new state as we increase the electron-phonon coupling is   of the first order~\cite{emil}. $C_\mathrm{el}=0$ marks the point where the metal  becomes unstable with respect to small deviations from the thermal equilibrium, i.e., ceases to be a local minimum of the free energy.  In first order phase transitions,  such a local (absolute) instability is preceded by a metastable region $\xi_c < \xi < \xi_*=1$.      The metal is no longer the global minimum of the free energy past  the phase-transition point $\xi_c$. Therefore, $\xi_*=1$ provides a fundamental upper bound on $\xi$ in   metals and superconductors.  This implies $\xi_c <1$, but the exact value of $\xi_c$ is nonuniversal and can depend on, for example, the  lattice and electronic band structures, number of carries per lattice site etc. 
 
 We analyzed data for numerous metals and superhydrides for which the Eliashberg functions are known and found\footnote{Code for computing $\xi$ from $\alpha^2F(\omega)$ is available at \href{https://doi.org/10.5281/zenodo.15550102}{https://doi.org/10.5281/zenodo.15550102}.} that all of them conform to our upper bound $\xi<1$.     Hydrides  loose stability at $\xi \le 0.5$, while  $\xi$ for other metals   (\textbf{Figure~\ref{figS1}} and Table~\ref{sc_table}) is significantly  lower, suggesting that $\xi_c=0.5$ for hydrides and $\xi_c< 0.5$ for conventional metals.  We attribute $\xi=0.58$   in CaH$_6$ and MgH$_6$ to the fact that these are not actually existing hydrides at these pressures. Moreover, if our conjecture that $\xi_c\le0.5$ holds, these two materials can only exist in a metastable state. 
 
  The condition $\xi<\xi_*=1$ implies also an upper bound $\lam<\lam_*$ on the standard interaction parameter $\lam$ defined as 
 \beg
 \lam=\int_0^\infty d\omega \frac{2\alpha^2F(\omega)}{\omega}= \sum_{k=1}^N \lam_k.
 \label{lam_def}
 \en
  The value of $\lam_*$ varies with the shape of the Eliashberg function  with a minimum  $\lam_*= 3.69$ attained for Einstein phonons (Section~\ref{sec:derivationTc}).  For   Debye phonons  ($\alpha^2 F\propto \omega^2$ for $\omega\le \omega_D$ and zero otherwise),  $\lam_*=4.72$.   For the Eliashberg function of the same shape  as that for a material with given 
  $\xi$ and $\lam$ (i.e., differing from it only by an overall scale), we have
 \beg
\lam_*=\frac{\lam}{\xi},\quad \lam_c=\frac{\lam \xi_c}{\xi}=\lam_*\xi_c.
\label{shape}
\en
In particular,  we obtain $\lam_*=7.39$ and (assuming $\xi_c=0.5$)  $\lam_c=3.69$  for YH$_9$ at 205 GPa (see Table~\ref{sc_table}) and $\lam_c=2.36$ for Debye phonons.  The values of $\lam_c$ for all experimentally realized hydrides are in the range from 2.19 for YH$_6$ to 4.12 for (La,Y)H$_{10}$.  This  indicates that  stability must be the reason why $\lam$ in actually existing metals is limited  to $\lam <4$.
  
  We argued previously that in the new state the system lowers its energy by either opening a gap or, at least, lowering the density of states at the Fermi energy~\cite{emil}. We also saw that the superconducting state is more resilient against perturbations than the normal state as it already has a gap. Therefore, we expect  robust  metastable superconductivity when $\lam$ is  quenched from $\lam<\lam_c$ to $\lam>\lam_c$ while the system is  superconducting. This  creates an opportunity of attaining  \textsl{metastable} superconductivity with substantially  higher $T_c$ in polyhydrides  at  lower pressures than usual, since $\lam$ generally increases with decreasing $P$, see \textbf{Figure~\ref{fig1}}. Moreover, metastable superconductivity of this type has likely    been observed in pressure-quenched  FeSe~\cite{fese}, but a more thorough analysis of these experiments is necessary to establish this with certainty.

 \section{Fundamental limit on   $T_c$ in phonon-mediated superconductors}
 
   In the formal limit $\lam\to\infty$ of the ME theory, $T_c$ asymptotically approaches $0.18 \sqrt{\lam \langle \omega^2\rangle}$, while at finite $\lam$ it falls below this asymptote~\cite{mitrovic,allen,michael}, i.e.,
 \beg
 T_c < 0.18 \sqrt{\lam \langle \omega^2\rangle}.
 \label{asym}
 \en
 This provides an upper bound on $T_c$  in terms of $\lam$ and the average square frequency~\cite{note_emp} 
 \beg
  \langle \omega^2\rangle = \frac{2}{\lam} \int_0^\infty \alpha^2 F(\omega)\omega d\omega.
 \en
 The requirement  $C_\mathrm{el}>0$ imposes a constraint $\lam<\lam_*$ on $\lam$. Let $\omega_{\max}$ be the maximum   available phonon frequency.  Spreading out the phonon spectral function to frequencies below 
 $\omega_{\max}$ increases $\lam_*$, while decreasing $ \langle \omega^2\rangle$ and results in an overall decrease of $T_c$ (see Section~\ref{sec:derivationTc}).  Therefore, the critical temperature    for  Einstein phonons ($T_c^E$) with maximal  $\lam=\lam_*=3.69$   and frequency $\omega_{\max}$  
 provides a \textsl{rigorous} upper bound on $T_c$  for  stable and metastable superconducting metals,
 \beg
 T_c < 0.32\, \omega_{\max},
 \label{tcmax}
 \en
 where we used a standard algorithm~\cite{allen,diag_algorithm} to  numerically compute $T_c^E=0.3175\, \omega_{\max}$.     We define a metal as a good conductor with a Fermi energy $E_F$ much larger than typical phonon frequencies. This is the only essential assumption that goes into the Eliashberg theory from which we derived \eref{tcmax}.    
 
 Recall that metals are stable for $\lam<\lam_c$ and metastable for $\lam_c<\lam<\lam_*$.  Assuming $\xi_c=1/2$ (see above),  \eref{shape} obtains $\lam_c=\lam_*/2$. Similarly calculating $T_c^E=0.1995\, \omega_{\max}$ for  Einstein phonons with  $\lam =3.69\slash2$, we
  determine an upper bound on $T_c$ for stable metals,
 \beg
 T_c < 0.20\, \omega_{\max}.
 \label{tcmax1}
 \en
 Metals with $0.20\, \omega_{\max} < T_c <0.32\, \omega_{\max}$ can therefore only exist in a metastable state.

 In the Debye model $\omega_{\max}$ is equal to the Debye frequency $\omega_D$. However, since real solids do not  conform  to the Debye model,  $\omega_D$ is not uniquely defined.  In particular,  its value  depends on the quantity used to extract it (e.g., the phonon specific heat vs resistivity).  We therefore prefer to use the edge of the phonon spectrum (Eliashberg function) as $\omega_{\max}$ whenever possible (Table~\ref{sc_table}). Note that both our upper bounds~\re{tcmax} and \re{tcmax1} are significantly
more generous than the heuristic  bound of $0.1\omega_D$ proposed in \cite{kivelson1}.   However, observe  that, for example, (La,Ce)H$_{9-10}$ at 123 GPa  with $T_c=190$ K  and $\omega_D=1107$ K, extracted from the temperature dependence of the electrical resistance,   violates the latter bound, so it apparently does not hold for compressed hydrides. Indeed, to fall below the $0.1\omega_D$ bound,  (La,Ce)H$_{9-10}$ must have a Debye frequency of about 2000 K, which exceeds all calculated and experimental values of $\omega_D$ ever obtained for synthesized superhydrides (Table~\ref{table_D}).

Our results suggest that room-temperature superconductivity is achievable in metals with  
$\omega_\mathrm{max} > 1500$~K and $\lam\ge 2$, which up until now has only been observed in hydrides at high pressure.   Moreover, since we have condensed the entire phonon weight to $\omega=\omega_{\max}$ to maximize $T_c$, the maximum  of the Eliashberg function cannot be far below $\omega=1500$~K in a room-temperature superconductor. 
 
  We can obtain an absolute numerical upper bound on $T_c$  by observing that the maximum phonon frequency cannot exceed the \textsl{ionic} plasma frequency and that interactions will only renormalize  $\omega_{\max}$ down~\cite{plasma}. The ionic plasma frequency  is inversely proportional to the square root of the ionic mass, and we therefore expect it to be the highest in the metallic hydrogen. According to recent  ab-initio calculations of the Eliashberg function for solid atomic hydrogen at 500 GPa~\cite{dangic}, $\omega_{\max}=3000\,\mathrm{cm}^{-1}=4320$ K for harmonic phonons (as the ones in the Eliashberg theory), see also~\cite{omega_max}. However, this Eliashberg function has a sharp maximum  below  $\omega_0=3000$ K dropping very quickly  from the maximum  to zero at  $\omega_{\max}$. In addition,  theoretical estimates of the Debye frequency for the metallic hydrogen  are in the range 3000--3500 K~\cite{ashcroft,ginzburg}.  It  is therefore     safe to replace $\omega_{\max}$  in \esref{tcmax} and \re{tcmax1} with $\omega_0=3000$ K, and we obtain the following absolute upper bounds on $T_c$ in stable and metastable metals:
  \beg
  T_c^\mathrm{stable} < \frac{600\,\mathrm{K}}{\sqrt{A}}, \quad T_c^\mathrm{metastable} < \frac{950\,\mathrm{K}}{\sqrt{A}},
  \label{ub}
  \en
 where $A$ is the atomic mass (in atomic units) of the lightest element in the material.

 It is important to note that $\omega_{\max}$ grows (roughly linearly) with pressure, see, e.g.,  \cite{omega_max} and   Figure~\ref{figure_supp}\blue{a}. For the above bounds, we used $\omega_{\max}$ at 500 GPa---the current upper limit for transport measurements. However, there is a negative correlation between $\omega_{\max}$ and other characteristic phonon frequencies   and the electron-phonon interaction strength  (Figure~\ref{fig1} and Figure~\ref{figure_supp}\blue{b}), which we did not take into account in our analysis, such that  increasing the pressure beyond 500 GPa will likely result in an overall decrease rather than increase of $T_c$.

 \section{Conclusion}\label{sec12}
 
 We  established  intrinsic upper bounds on the electron-phonon interaction strength and superconducting $T_c$  in   metals. Materials where this interaction exceeds a certain threshold cannot exist in nature in a metallic state, similar to the states  of the Van der Waals gas with negative compressibility.    Just as in the case of the Van der Waals  gas, there is an absolute instability (negative electronic specific heat)  that signals a first order phase transition without immediately  telling us what the new phase is.  However, based on additional considerations~\cite{emil} and  experiment, we believe that the tendency is   towards a lattice reconstruction.

 It is generally known that superconducting $T_c=\kappa \omega_{\max}$, where $\omega_{\max}$ is the maximum or  some other characteristic phonon frequency and the coefficient $\kappa$ is a monotonically increasing function of the strength of the electron-phonon interaction (overall height of the Eliashberg function). As a result, a fundamental bound on $T_c$  is impossible without a limit on  the interaction strength. Prior work suggested that $\kappa$ might be somehow limited by stability, but was unsuccessful in furnishing definitive evidence  of an instability as well as in determining its character and  criteria for it (see also Section~\ref{sec:polaronic}).   Our work fills this crucial gap and provides the precise stability limit. 
 
 It is clear from the bounds~\re{ub}   that stable room-temperature phonon-mediated superconductivity can only be achieved in  hydrogen $(A=1)$ and deuterium $(A=2)$ compounds. Substitution of hydrogen with deuterium typically decreases $T_c$ by a factor of roughly 1.4~\cite{drozdov5,kong,chen,drozdov1,troyan1,isotope1} (isotope effect) consistent with \eref{ub}.  Helium-3 and 4 are extremely unlikely candidates~\cite{helium}, and, in addition, $T_c<300$ K already for $A=4$.

 At the same time, there are no fundamental reasons why $T_c$ cannot exceed room temperature in hydrides, at least under sufficiently high pressure. All that is required to engineer such superconductors is an electron-phonon coupling constant $\lambda = 2$ -- $3$, arising primarily from phonons with frequencies near or above 1500~K.   Moreover, we showed that the upper bound on $T_c$ increases by a factor of 1.59 for metastable metals, and that further enhancement is possible if the pressure (and hence the electron-phonon coupling) is quenched while the material is in the superconducting state, as observed in recent FeSe experiments~\cite{fese}.

 Our results provide a theoretical foundation for the pursuit of room-temperature superconductivity in polyhydrides, demonstrating that such temperatures are not precluded by any fundamental limits.
 This conclusion is supported by recent first-principles predictions of high-$T_c$ superconductivity in binary and ternary hydride compounds such as ScH$_{12}$~\cite{jiang2024adv}, LaSc$_2$H$_{24}$~\cite{he2024pnas}, and related systems. On the experimental side, although progress is more gradual, mounting evidence points to the emergence of new superconducting phases such as LaH$_{12}$~\cite{semenok2024arxiv} within the extensively studied La--H~\cite{somayazulu2019prl,wu2024natcomm,struzhkin2020mre} and ternary La--Sc--H~\cite{semenok2025afm} systems. These materials occasionally exhibit partial transitions in the range of 265--280~K, already approaching room temperature. While experimental challenges persist---including the synthesis of pure phases, precise control of stoichiometry, and stabilization under lower pressures---these advances highlight a promising route toward room-temperature superconductivity in hydrides.
 
 \bigskip
 
\textbf{Acknowledgements}   

\medskip

 Dmitri V. Semenok's research was supported  in part  by the National Natural Science Foundation of China (NSFC, grant No. 1231101238) and Beijing Natural Science Foundation (grant No. IS23017).
 
 \bigskip

  \textbf{Author contributions}
  
  \medskip
  
D.V.S. collected statistical data, prepared  figures and tables, and performed numerical calculations.  E.A.Y. and B.L.A. contributed  theoretical analysis. E.A.Y. carried out  analytical and related numerical calculations. E.A.Y. and D.V.S. wrote the text.  All authors discussed  the results and made helpful inputs to all parts of the manuscript.

\bigskip
 
  \textbf{Competing interests}
  
  \medskip
  
 The authors declare no competing interests.
 
 \bigskip
 
  \textbf{Data Availability Statement} 
  
  \medskip
  
  The data that support the findings of this study are available from the  authors upon reasonable request. The code for computing the stability parameter $\xi$ from the Eliashberg function $\alpha^2F(\omega)$ is available at \href{https://doi.org/10.5281/zenodo.15550102}{https://doi.org/10.5281/zenodo.15550102}.

\bigskip
 
  \textbf{Keywords} 
  
  \medskip
  
 electron-phonon interaction, high-pressure, hydrides, metals, superconductivity

\bigskip


 
\begin{center}
\textbf{\LARGE Supporting Information}
\end{center}

\setcounter{equation}{0}
\setcounter{figure}{0}
\setcounter{table}{0}
\newcounter{mycounter}[part] 
\renewcommand{\thesection}{S\arabic{mycounter}} 
\makeatletter
\renewcommand{\theequation}{S\arabic{equation}}
\renewcommand{\thefigure}{S\arabic{figure}}
\renewcommand{\thetable}{S\arabic{table}}
\renewcommand{\theHtable}{S\thetable}
\renewcommand{\theHfigure}{S\thefigure}
\renewcommand{\theHequation}{S\theequation}

\addtocounter{mycounter}{1}


\section{Stability analysis}
\label{sec:stability_analysis}

\addtocounter{mycounter}{1}

To evaluate the stability of a metal with respect to  electron-phonon interactions, we use the standard kinetic equation for the  electron distribution function $f(E, t)$~\cite{prange1,rammer},
\beg
\left(1-\frac{\partial\Sigma}{\partial E}\right)\frac{\partial f}{\partial t}+ \frac{\partial\Sigma}{\partial t} \frac{\partial f}{\partial E}=
a_\mathrm{ep} I_\mathrm{ep}(E)+a_\mathrm{ee} I_\mathrm{ee}(E),
\label{kinetic}
\en
where we added nonnegative constant coefficients $a_\mathrm{eph}$ and $a_\mathrm{ee}$ for convenience.  The electron self-energy is
\beg
\Sigma=\fint dE'\int_0^\infty\!\!\! d\omega\, \alpha^2 F(\omega) \frac{ f(E'+\omega) - f(E'-\omega)}{E-E'},
\en
and the electron-phonon collision integral is 
\beg
I_\mathrm{ep}(E)=-2\pi \int_0^\infty\!\!\! d\omega\, \alpha^2 F(\omega) \left\{ N_0(T_\mathrm{ph}) [2f-f_+-f_-] +f(f_+-f_-)+f-f_+\right\},
\en
where $f\equiv f(E,t)$, $f_\pm\equiv f(E\pm\omega,t)$, and 
\beg
N_0(T_\mathrm{ph})=\frac{1}{e^{\omega/T_\mathrm{ph}}-1}
\en
 is the equilibrium Bose (phonon) distribution at temperature $T_\mathrm{ph}$. The kinetic equation~\re{kinetic} assumes  spatially uniform initial conditions and, as usual, that phonons remain in thermal equilibrium, see below and, e.g., \cite{rammer}, which also provides an explicit expression for the electron-electron collision integral $I_\mathrm{ee}.$ 

 We will prove that the metal is unstable when $C_\mathrm{el}<0$ by contradiction.  Let 
\beg
f(E, t=0)=f_0(T)\equiv \frac{1}{e^{E/T}+1}
\label{ini}
\en
 be the equilibrium Fermi distribution  with temperature $T$ that is slightly higher than the phonon temperature $T_\mathrm{ph}$. Experimentally, initial conditions of this type are created by heating the electrons with an ultrashort laser pulse~\cite{optical1,optical2,optical3}.  Note, however, that we do not assume the two-temperature model, but merely
 chose an initial condition of the form~\re{ini}. The true fermion distribution   $f(E,t)$ will generally become nonthermal   in the course of the actual time evolution of the system.

 Suppose the system is stable with respect to the electron-phonon interaction ($a_\mathrm{ep}=1$, $a_\mathrm{ee}=0$).  Since the phonon specific heat is much larger than that of the electrons, $C_\mathrm{ph}/|C_\mathrm{el}|\sim E_F/\omega_D\gg1$, the change in the phonon temperature and, more generally, the deviation of the phonon distribution   from the equilibrium Bose distribution $N_0(T_\mathrm{ph})$ is negligible. Phonons serve as a thermal bath for the electrons, and electrons equilibrate at temperature $T_\mathrm{ph}$ up to corrections of order $\omega_D/E_F$, which are beyond the accuracy of the Eliashberg theory anyway.  

Linearizing the kinetic equation and $f(E,t)$ around the equilibrium at $T=T_\mathrm{ph}$ with the help of the usual substitution~\cite{kinetics},
\beg
f(E,t)=f_0+\delta f\equiv f_0(T_\mathrm{ph}) + \frac{f_0(T_\mathrm{ph})\left[1-f_0(T_\mathrm{ph})\right]}{T_\mathrm{ph}} \varphi(E,t),
\en
 we obtain after some algebra
\beg
\int \!\! dE' A(E,E') \dot\varphi(E', t) =-\int \!\! dE' \left[ a_\mathrm{ep}  M_\mathrm{ep} (E, E') + a_\mathrm{ee} M_\mathrm{ee} (E, E') \right] \varphi(E', t),
\en
where $A(E, E')$, $M_\mathrm{ep} (E, E')$, and $M_\mathrm{ee} (E, E')$ are \textsl{real symmetric} integration kernels (matrices) and $\dot\varphi\equiv \partial\varphi/\partial t$.  Letting $\varphi(E,t)=e^{-\gamma t}\psi(E)$, we reduce the problem to a generalized eigenvalue equation of the form $\gamma A \psi= M\psi$. It is straightforward to show that  $M_\mathrm{ep}$ and $M_\mathrm{ee}$ are positively defined for any electron-phonon couplings $\lam_i$ [by, e.g., considering  the weak coupling limit  where the system is obviously stable ($\gamma>0$) for any physical phonon spectrum]. The assumption that the system is stable with respect to the electron-phonon interaction (i.e., for $M= M_\mathrm{ep}$) implies that $A$ is positively defined as well. Since a  linear combination  of two positively defined real symmetric matrices with positive coefficients   is similarly positively defined, it follows that  it must also stable for $M=a_\mathrm{ep} M_\mathrm{ep}+ a_\mathrm{ee} M_\mathrm{ep}$ for any $a_\mathrm{ep}>0$ and $a_\mathrm{ee}>0$.

Therefore, if the metal is stable with respect to the electron-phonon interaction,  \eref{kinetic} must also be linearly stable for any choice of   $a_\mathrm{ep}>0$ and $a_\mathrm{ee}>0$. Let $a_\mathrm{ee}\gg a_\mathrm{ep}>0$. Then, electron-electron collisions dominate, and the electron distribution    thermalizes essentially instantaneously with temperature $T(t)$, i.e., $f(E, t)=f_0(T(t))$,  corresponding to the instantaneous energy density $\epsilon(t)$ of the electronic subsystem. However, since electron-electron collisions conserve $\epsilon(t)$ (this follows formally from $\int\! dE E I_\mathrm{ee}(E)=0$), the latter is able to relax only through electron-phonon collisions.  Multiplying  \eref{kinetic} by $E$, integrating over $E$, and using $f(E, t)=f_0(T(t))$, we obtain
\beg
C_\mathrm{el}\frac{ dT}{dt}=2\pi \nu_0 a_\mathrm{ep} \int_0^\infty \!\!\! d\omega\alpha^2 F(\omega)\omega^2\left[ N_0(T_\mathrm{ph})-N_0(T)\right],
\label{inst}
\en
where
\beg
C_\mathrm{el}=2\nu_0\int \!\! dE E \left\{\left(1-\frac{\partial\Sigma_0}{\partial E}\right)\frac{\partial f_0}{\partial T}+ \frac{\partial\Sigma_0}{\partial T} \frac{\partial f_0}{\partial E}\right\}
\label{PK}
\en
is the electronic specific heat. Here $\Sigma_0$ and $f_0$ are the equilibrium electron self-energy and distribution function at temperature $T$. \eref{PK}   for $C_\mathrm{el}$   was derived by Prange and Kadanoff in 1964~\cite{prange1} and later shown to be equivalent to \eref{C} in the main text~\cite{lee1,dolgov1}. Note that there is a typo in \cite{prange1}---the sign in front of $\frac{\partial\Sigma_0}{\partial E}$ should be minus and not plus as they have.

The right hand side of \eref{inst} is negative when $T>T_\mathrm{ph}$ indicating that the heat flows from hotter electrons to the colder phonon bath. Linearizing this equation in $(T-T_\mathrm{ph})$, we see that
$T$ grows exponentially when $C_\mathrm{el}<0$. Note that \eref{inst} does not describe the actual dynamics of the system but is a consequence of the assumption that it is stable with respect to the electron-phonon interaction. Thus, $C_\mathrm{el}<0$  is a \textsl{sufficient} condition of the instability. For further details on this stability analysis, see \cite{kinetic_methods}. This reference also demonstrates that the loss of kinetic stability occurs at smaller values of $\lambda$, while $C_\mathrm{el}$ remains positive---that is, before the equilibrium Migdal-Eliashberg theory detects instability.

For $\lam>\lam_*$, the electronic specific heat is negative in a temperature interval from $T_-$ to $T_+>T_c$. In particular, $T_-\to 0$ and $T_+\to 2.1335 T_c$ in the strong coupling limit $\lam\to\infty$~\cite{breakdown}. However,
a more complete kinetic analysis  shows that the   normal state remains unstable for all temperatures below $T_-$ as well, emphasizing that $C_\mathrm{el}<0$ is sufficient but not necessary for the instability. To put it another way, the normal state is unstable for all $T<T_+$ even though $C_\mathrm{el}<0$ only  for $T_-<T<T_+$. We also note that the existence of a new global minimum of the free energy just above $T_c$ implies by continuity that the superconducting state is metastable at least for temperatures not too much below $T_c$~\cite{breakdown}.

\section{Proof of the stability condition $\xi <1$}
\label{sec:proof}

\addtocounter{mycounter}{1}

 Since the primary theoretical result presented in this manuscript is the stability condition expressed by inequality \re{C>0} in the main text, we  explicitly outline the logic underlying its derivation here rather than relying on previous publications.

 ME theory is obtained from the stationary point of the action, describing electrons interacting through a retarded potential mediated by physical phonons. Provided the system is a stable metal, this stationary point corresponds to a global minimum of the thermodynamic potential~\cite{breakdown}. Corrections to this stationary solution, known as loop corrections, are controlled by a small parameter---the ratio of the maximum phonon frequency to the Fermi energy~\cite{agd_m,migdal_m,eli1st_m,1loop}. This fact, although presented differently, was originally demonstrated by Migdal and Eliashberg themselves~\cite{migdal_m,eli1st_m}, who observed lattice instability at a bare coupling $\lam_0\approx 0.5$ within the Fr\"ohlich Hamiltonian. They emphasized two essential conditions for their theory's validity: (1) a small ratio of maximum phonon frequency to Fermi energy, and (2) a stable lattice. This second condition has often been overlooked in subsequent studies employing effective electron-phonon models.

 In typical metals, the ratio of the maximum phonon frequency to the Fermi energy ranges between $10^{-2}$ and $10^{-4}$, thus providing excellent accuracy for evaluating the electronic specific heat. Indeed, it is well-established that the ME calculation of electronic specific heat consistently matches experimental results within a few percent accuracy~\cite{2008review,carbotte,weird,golubov}.

 {Our stability condition is proven by contradiction: assuming a stable metal described by a given Eliashberg function with Fermi energy much larger than the maximum phonon frequency, the ME formula for electronic specific heat, \eref{C} in the main text, must apply. However, the kinetic stability analysis of the previous subsection reveals instability when the specific heat becomes negative. Thus, a metal characterized by a negative ME electronic specific heat cannot physically exist, establishing inequality~\re{C>0}. %
 }

\section{Kinetic instability vs. polaronic and CDW instabilities arising from effective electron-phonon models}
\label{sec:polaronic}

\addtocounter{mycounter}{1}

It is important to clearly distinguish the kinetic instability, which yields our stability condition, \eref{C>0} in the main text, from the well-known polaronic\slash charge-density-wave (CDW)\slash lattice  instabilities  arising in effective electron-phonon models~\cite{agd_m,migdal_m,eli1st_m,Chakraverty,scalettar,div2,kabanov,roland,ono,alexandrov,capone,kresin,esterlis,scalapino,Nosarzewski,Bradley,ZSRBMPS}
{ (such as the Fr\"ohlich or Holstein Hamiltonians). A key feature of our kinetic instability is that it establishes a strict upper bound on the \textsl{physical} electron-phonon interaction strength $\lambda$---a bound that, to the best of our knowledge, has not been previously identified. In contrast, polaronic\slash CDW transitions do not impose a sharp upper limit on $\lambda$, as $\lambda \to \infty$ at the onset of such instabilities~\cite{agd_m,migdal_m,eli1st_m,scalettar,div2,valid}.

Secondly, while the polaronic\slash CDW\slash lattice instability describes an equilibrium phase transition, our kinetic instability is inherently a nonequilibrium phenomenon, occurring prior to any equilibrium phase formation.  Moreover, the subsequent evolution following our instability does not necessarily lead to polaronic or CDW phases. For example,  pressurized hydrides follow a completely different route, namely, hydrogen diffusion and the formation of new chemical compounds~\cite{kong_methods,chen_m,drozdov1_methods,minkov1_m,chen3_methods,48}.

Finally, the  lattice instability observed in effective electron-phonon models  is an artifact of these models, which at best, obscures the accompanying polaronic or CDW transitions. Fundamentally, the physical system is governed by the Coulomb Hamiltonian, describing electrons and ions interacting via Coulomb forces. Phonons naturally emerge as collective quasiparticle excitations from this Hamiltonian, inherently incorporating electron-ion interactions. Specifically, the electron density dynamically adjusts to ionic movements, leading to electronic polarization which directly shapes the phonon spectrum.

In simplified models such as the Holstein or Fr\"ohlich Hamiltonians, the phonon frequency renormalization improperly counts the electronic polarization twice. This double-counting issue results in artificial phonon softening or lattice instability at a \textsl{bare} electron-phonon coupling $\lambda_0 \approx 0.5$. This spurious instability arises from incorrectly including the static part of the electron polarization operator more than once~\cite{kagan,geilikman}, see also \cite{kresin,double3,screen,marini}.
Rigorous adiabatic perturbation theory starting from the Coulomb Hamiltonian does not produce such lattice instabilities~\cite{kagan,geilikman}. Hence, the instability seen in simplified electron-phonon models is essentially an artifact of double-counting (or ``overscreening''~\cite{screen,marini})---with ion-ion interactions effectively screened twice.

Indeed, Holstein, Fr\"ohlich, and related models cannot be consistently derived from the underlying Coulomb Hamiltonian at any order of adiabatic perturbation theory~\cite{geilikman}, resulting in fundamental limitations. These effective models successfully capture phonon-mediated electron-electron interactions but fail to provide accurate phonon spectra. In reality, the Coulomb Hamiltonian naturally produces a unique phonon spectrum. Therefore, the standard and widely accepted practice in phonon-mediated superconductivity is to treat the phonon spectrum as an external input rather than attempting to renormalize it within these simplified effective models~\cite{mitrovic,carbotte,2008review,Sadovskii_m}.

Our approach avoids these pitfalls by directly working with physical phonons, employing the Eliashberg function as a fixed input. Rather than relying on effective electron-phonon models, our theory addresses electrons interacting through retarded, phonon-mediated interactions. Consequently, a consistent formulation of our theory is inherently non-Hamiltonian, expressed instead via a nonlocal fermionic action, with the nonlocality reflecting the retarded nature of interactions mediated by physical phonons.

\section{Derivation of upper bounds on the electron-phonon interaction constant $\lam$ and $T_c$}
\label{sec:derivationTc}

\addtocounter{mycounter}{1}

 It is helpful to disentangle the notions of the shape,  height (overall scale), and  support (spread) of the Eliashberg function $\alpha^2F(\omega)$ from each other. To this end, we introduce a normalized \textsl{shape} function
\beg
P(\omega)=\frac{1}{\lam}\frac{2\alpha^2 F(\omega)}{\omega}.
\en 
By the definition of $\lam$ in \eref{lam_def} in the main text and since $\alpha^2F(\omega)\ge0$,
\beg
\int_0^\infty\!\!\! P(\omega) d\omega=1,\quad P(\omega)\ge 0,
\en
so we can think of $P(\omega)$ as the distribution function of phonon frequencies. It determines the shape of the Eliashberg function, while   $\lam$ controls its overall height. Note that we can specify $\lam$ and $P(\omega)$ independently of each other.

 Now it is straightforward to determine the upper bound $\lam_*$ on the electron-phonon coupling $\lam$ from the stability condition [\eref{C>0} in the main text],
 \beg
\lam_*=\frac{1}{\max_T \left\{\int_0^\infty   g \left( \frac{\omega}{2\pi T}\right)  P(\omega) d\omega \right\}}.
\label{lam*}
\en
The function $g(x)$ reaches its single maximum $g_{\max}=0.2709$ at $x_{\max}= 0.3273$. The upper bound $\lam_*$ is smallest, $\lam_*=1/g_{\max}=3.6915$, for  Einstein phonons, i.e., for $P(\omega)=\delta(\omega-\omega_E)$, because   in this case the denominator of \eref{lam*} assumes its maximum possible value $g_{\max}$.  Similarly, \eref{lam*} provides $\lam_*$ for  other phonon spectra, e.g., for the Debye model [$P(\omega)=2\omega/\omega_D^2$ for $\omega<\omega_D$ and zero otherwise] or for the same spectrum [same $P(\omega)$] as that for any of the materials in   Table~\ref{sc_table}. Furthermore,
the definition of the stability parameter $\xi$ in \eref{C>0} in the main text together with \eref{lam*} imply \eref{shape} in the main text.

To bound $T_c$, we need to bound both $\lam$ and the phonon spectrum. Suppose the highest phonon frequency is $\omega_{\max}$ and let $x=\omega/\omega_{\max}$. Then,
\beg
 \lam \langle \omega^2\rangle =\omega_{\max}^2  \frac{\int_0^1 x^2 P(x) dx}{\max_\tau \left\{\int_0^1   g \left( \frac{x}{\tau}\right)  P(x) dx \right\}}.
\en
By analyzing  the variational derivative of the right hand side of this equation with respect to $P(x)$, we find that it is maximal for $P(x)=\delta(x-1)$, i.e., for  Einstein phonons with frequency $\omega_E=\omega_{\max}$. Taking this into account,
we immediately obtain \eref{tcmax} from \eref{asym} in the main text. 



\section{Properties and evolution of superconducting materials}

\addtocounter{mycounter}{1}

\begin{figure}[H]
\centerline{\includegraphics[width=0.92\columnwidth]{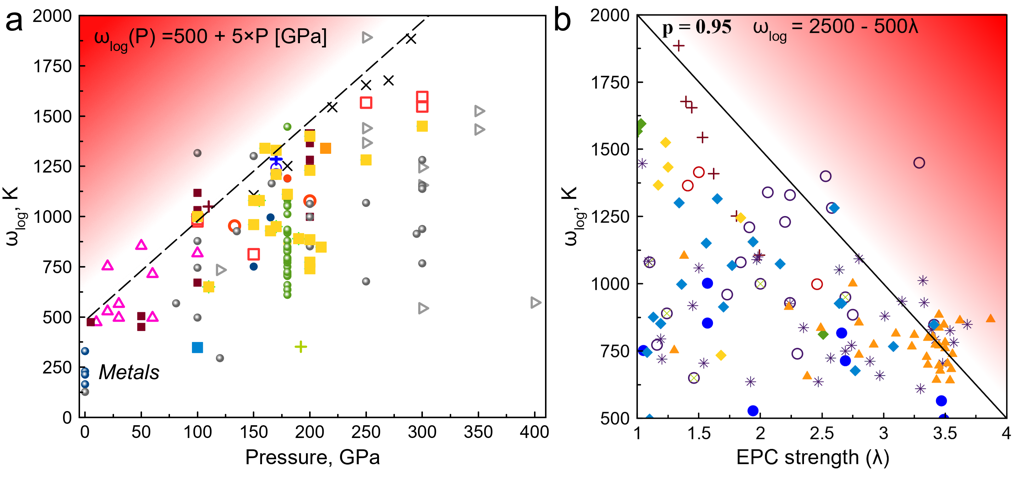}}
\caption{Statistical analysis of the results of   DFT calculations of electron-phonon interaction parameters in metals, polyhydrides, and other compounds. (a) Correlation between the logarithmic average phonon frequency $\omega_\mathrm{log}$ in Kelvin and pressure $P$ in GPa. For about 95\% of the  compounds $\omega_\mathrm{log}$    lies below the line ${500 + 5P}$. (b) Correlation between  $\omega_\mathrm{log}$ in Kelvin and the electron-phonon coupling parameter $\lam$. For about 95\% of the  compounds  $\omega_\mathrm{log}$ in Kelvin  falls below the line ${2500-500\lam}$. Red areas  indicate unlikely combinations of $\omega_\mathrm{log}$, $P$ and $\lam$. Data for the plot  are from Refs.~\cite{12,13,22,49,50,51,52,53}.}
\label{figure_supp} 
\end{figure}

 \begin{table}
\centering
 \caption{Various superconducting metals, compounds, and compressed polyhydrides and the values of $T_c,$ electron-phonon coupling constant  $\lam$,  stability parameter   $\xi$,  and average logarithmic ($\omega_\mathrm{log}$) and  maximum ($\omega_\mathrm{max}$)  phonon frequencies  for them.    }
\begin{tabular}{cccccc}
\toprule
 \textbf{Compound} & \textbf{$\bm T_c$, K} & $\bm \lam$  &  $\bm \xi$  &  \makecell{$\bm\omega_\mathrm{\textbf{log}},$  \textbf{K} } & $\bm \omega_{\bm\max},$ \textbf{K}  \\ 
\midrule 
\multicolumn{6}{c}{\textsl{Experimental Materials}}\\
\\
Ga (amorph.)\cite{33} &	8.6 &	2.25 & 0.16 &	62	& 302     \\
Pb (amorph.)\cite{33}  &	7.2	& 1.91 & 0.29 & 35 &155    \\
 
Bi (amorph.)\cite{33} 	& 6.1	 &1.84 -- 2.46 & 0.19 &	42 &163	    \\
 
Hg\cite{33}  &	4.2	& 1.0 -- 1.6 & 0.14 &	 86	& 165    \\
 
Nb\cite{33} 	&9.2&	0.82 -- 1.05 & 0.2 &	 163	& 325    \\
 
Nb$_3$Sn\cite{stewart} &	17.9	& 1.6 -- 1.8& 0.27 &	142	& 422   \\
 
MgB$_2$\cite{kong2}  &	40	&0.87 &0.16&	680  & 1624     \\
 
PbBi (amorph.)\cite{chen2} &	7.0	& 3.0	 &0.26& 33.3   &  150     \\
 
PbBi3 (amorph.)\cite{chen2}  &	6.8	& 2.78 &0.23& 	33.4  & 154    \\
 
H$_3$S (157 GPa)\cite{drozdov1_methods} &	190	& 1.84 &0.32&	1078   & 2704     \\
 
LaH$_{10}$  (214 GPa)\cite{drozdov5_methods,Somayazulu_methods} &	245 &	2.06 & 0.43 &	1340  & 3145   \\
 
YH$_6$ (165 GPa)\cite{40} &	224	& 1.71 & 0.39 &	1333   & 2450  \\
 
ThH$_9$ (150 GPa)\cite{dima1}	& 146	& 1.73	& 0.32 & 957  & 3362    \\
 
ThH$_{10}$ (170 GPa)\cite{dima1,Kvashnin}	& 161	& 1.65 & 0.31 &	1116 -- 1470 & 3122  	 \\
 
YH$_9$ (205 GPa)\cite{kong_methods,12}	& 235 &	2.66	&0.36& 916 -- 1054 & 3167   	 \\
 
(La,Y)H$_{10}$ (180 GPa)\cite{dima_methods} &	253	& 3.87 & 0.47 &	868  &  3119   \\
 
(La,Ce)H$_9$ (123 GPa)\cite{chen3_methods,bi}	& 190 &	2.27	& 0.33 & 582 -- 915 & 2894    \\

\midrule
\multicolumn{6}{c}{\textsl{DFT Calculations}}\\
\\
CaH$_6$ (172 GPa)\cite{feng}	 & 215	& 2.69 & 0.58\tnote{*} &	950  &  2877   \\
LaH$_{16}$ (250 GPa)\cite{kruglov1} & 141 & 1.89 & 0.31 & 1511 & 3481   \\

ScH$_{12}$ (200 GPa)\cite{jianga} & 325 &  2.85 & 0.47 & 1189 & 2959    \\
 
Li$_2$MgH$_{16}$ (250 GPa)\cite{sun-lv}	 & 473 & 3.30 & 0.49 & 1111 & 3783   \\
 
MgH$_6$\cite{feng} & 263 & 3.29 & 0.58 & 1408 & 3719    \\
 
Hydrogen (\textsl{I}41/\textsl{amd}, 500 GPa)\cite{dangic_methods} & 374 & 2.85 & 0.45 & 1616 & 4089  \\
\bottomrule
\end{tabular}
\label{sc_table} 
\end{table}

\begin{table}
\centering
\caption{Data for Figure~\ref{fig0}:  evolution of superconducting materials in  20th -- 21st centuries.}
\begin{tabular}{lcc}
\toprule
\textbf{Material}     & \textbf{Year of discovery} & \textbf{Critical temperature, K} \\ \midrule
\multicolumn{3}{c}{\textsl{Low-temperature superconductors\cite{9}} }         \\ \\
Hg            & 1911              & 4.2                     \\
Pb            & 1914              & 7.3                     \\
Nb            & 1933              & 9.7                     \\
NbN           & 1941              & 16                      \\
Nb$_3$Sn         & 1952              & 18.3                    \\
V$_3$Si          & 1953              & 17.1                    \\
Nb$_3$Ge         & 1973              & 23.2                    \\
NbTi          & 1962              & 9.2                     \\
CaC$_6$\cite{10}          & 2005              & 11.5                    \\
MgB$_2$          & 2001              & 40                      \\  \midrule
\multicolumn{3}{c}{\textsl{Unconventional superconductors\cite{11}}}          \\ \\
KWO$_3$          & 1967              & 6                       \\
LiTi$_2$O$_4$       & 1973              & 1.2                     \\
BaPbBiO$_3$      & 1975              & 13                      \\
La$_2$BaCuO$_4$     & 1986              & 30                      \\
YBa$_2$Cu$_3$O$_7$     & 1987              & 90                      \\
BaKBiO$_3$       & 1988              & 20                      \\
BiSrCaCu$_2$O$_6$   & 1988              & 105                     \\
TlBa$_2$Ca$_2$Cu$_3$O$_9$ & 1989              & 110                     \\
HgBa2CaCu2O6  & 1993              & 120                     \\
GdFeAsO       & 2008              & 53.5                    \\
SrFFeAs       & 2009              & 72                      \\ \midrule
 &\textsl{Superhydrides\cite{9,12}} &                          \\ \\
LaH$_{10}$         & 2018              & 250                     \\
ThH$_9$          & 2020              & 146                     \\
ThH$_{10}$         & 2020              & 161                     \\
YH$_6$           & 2021              & 224                     \\
YH$_9$           & 2021              & 243                     \\
CeH$_9$          & 2021              & 100                     \\
CeH$_{10}$         & 2021              & 115                     \\
BaH$_{12}$         & 2021              & 20                      \\
YH$_4$           & 2022              & 82                      \\
(La,Y)H$_{10}$     & 2021              & 253                     \\
Lu$_4$H$_{23}$        & 2023              & 70                      \\
SnH$_4$          & 2022              & 74                      \\
PH$_3$           & 2015              & 100                     \\
SbH$_4$          & 2023              & 115                     \\
(La,Ce)H$_{9-10}$  & 2023              & 190                     \\
NbH$_3$          & 2024              & 42\\  \bottomrule                  
\end{tabular}
\end{table}

\section{Parameters of the electron-phonon interaction in various metals, alloys, and superhydrides}

\addtocounter{mycounter}{1}
 
\begin{table}[H]
\caption{Parameters of the electron-phonon interaction in metals, polyhydrides and, other compounds. We used this data  to draw Figure~\ref{fig1}. }
\centering
\begin{tabular}{cccc}
\toprule
\textbf{Compound} & \textbf{Pressure, GPa} & \textbf{EPC strength ($\bm \lam$)} & $\bm \omega_\mathrm{log}$\textbf{, K}\\
\midrule
\multicolumn{4}{c}{\textsl{DFT predictions}}\\
ScH$_9$~\cite{13} & 300 & 1.94 & 1156\\
VH$_8$~\cite{14,15} &  200 & 1.13 & 876\\
CrH$_3$~\cite{16} &  81 & 0.95 & 568\\
ZrH$_6$~\cite{17,18} & 295 & 1.7 & 914\\
ZrH$_{10}$~\cite{19} & 250 & 1.77 & 1068\\
NbH$_4$~\cite{20,21} &  300 & 0.82 & 938\\
SrH$_6$~\cite{22} & 100 & 1.65 & 1316\\
HfH$_{10}$~\cite{19} & 250 & 2.77 & 677\\
TaH$_6$~\cite{23} & 300 & 1.56 & 1151\\
TiH$_{14}$~\cite{22,24} & 200 & 0.81 & 1063\\
Hf$_3$H$_{13}$~\cite{22} & 100 & 1.1 & 497\\
RaH$_{12}$~\cite{22} &  200 & 1.36 & 998\\
MgH$_6$~\cite{feng} &  300 & 3.29 & 1450\\
CaYH$_{12}$~\cite{26} &  200 & 2.2 & 1230\\
MgCaH$_{12}$~\cite{27} & 200 & 2.53 & 1400\\
YH$_{10}$~\cite{28,29} & 250 & 2.58 & 1282\\
ScH$_{12}$~\cite{30} & 200 & 2.85 & 1189\\
Li$_2$MgH$_{16}$~\cite{sun-lv} & 250 & 3.3 & 1111\\
ScH$_4$~\cite{13, 32} & 250 & 0.81 & 1892\\
\midrule
\multicolumn{4}{c}{\textsl{Metals and alloys (experiment)}}\\
Nb~\cite{33} & 0 &1.05 &229\\
V~\cite{33} &  0 & 0.83 & 330\\
Sn~\cite{33} & 0 & 0.72 & 165\\
Ta~\cite{33} & 0 & 0.73 & 213\\
Hg~\cite{33} & 0 & 1.3 & 86\\
a-Ga~\cite{33} & 0 & 2.25 & 62\\
a-Pb{*}\tnote{*}~\cite{33} & 0 & 1.91 & 35\\
a-Bi~\cite{33} & 0 & 2.46 & 42\\
Nb$_3$Sn~\cite{33} & 0 & 1.7 & 142\\
MgB$_2${**}\tnote{**}~\cite{kong2} & 0 & 0.87 & 680\\
a-PbBi~\cite{chen2} & 0 & 3.0 & 33.3\\
a-PbBi3~\cite{chen2} & 0 & 2.78 & 33.4\\
\midrule
\multicolumn{4}{c}{\textsl{Hydrides (experiment)}}\\
(La,Y)H$_{10}$~\cite{dima_methods} & 180 & 3.87 & 868\\
(La,Ce)H$_{9-10}$~\cite{chen3_methods} & 123 & 2.27 & 915\\
H$_3$S~\cite{38} & 157 & 1.84 &1080\\
LaH$_{10}$~\cite{39} & 163 & 2.67 &1118\\
YH$_9$~\cite{12} & 200 & 2.75 & 885\\
YH$_6$~\cite{40} & 170 & 2.24 & 1330\\
ThH$_{10}$~\cite{dima1} & 170 & 1.91 & 1210\\
ThH$_9$~\cite{dima1} & 150 & 1.73 & 960\\
YH$_4$~\cite{40} & 155 & 1.1 & 1080\\
CeH$_9$~\cite{42} & 110 & 1.46 & 650\\
CeH$_{10}$~\cite{43} & 100 & 2.0 & 1000\\
SnH$_4$~\cite{44} & 190 & 1.24 & 890\\
CaH$_6$~\cite{45} & 170 & 2.69 &950\\ 
                         \bottomrule
\end{tabular}
\begin{tablenotes}
\item[*]*Amorphous film of Pb
\item[**]**Given for comparison
\end{tablenotes}
\end{table}

\begin{table}

\caption{Values of the stability parameter $\xi,$ electron-phonon coupling constant $\lam$, and the average logarithmic frequency $\omega_\mathrm{log}$ for several hydrides at different pressures $P$.}
\centering
\begin{tabular}{ccccc}
\toprule
\textbf{Compound}               & $\bm P$, \textbf{GPa} & $\bm \xi$     & $\bm\lam$ & $\bm  \omega_\mathrm{log},$ \textbf{K}                                                                                                           \\
\midrule
\multirow{4}{*}{CeH$_9$}  & 100           & 0.175 &  &\\  
                       & 120           & 0.165 &  &   \\ 
                       & 150           & 0.152 & &    \\ 
                       & 200           & 0.131 &  &    \\
                         \midrule
\multirow{5}{*}{LaH$_{10}$} & $\sim$50*\tnote{*}     & 0.96  &  5.32 &   \\                                                                                                                          
                       & 129           & 0.64  &  3.62 &   887                              \\
                       & 163           & 0.53  &     2.67 &1119                              \\
                       & 214           & 0.43  &     2.06 & 1340                            \\ 
                       & 264           & 0.37  &  1.73 &1469       \\   
                       \midrule
\multirow{2}{*}{H$_3$S}   & 135           & 0.42  &  &       \\
                       & 157           & 0.39  &  &     \\ 
                       \midrule
\multirow{3}{*}{ThH$_{10}$} & 100           & 0.46  &  2.57 & 1048      \\
                       & 200           & 0.32  &   1.58 & 1184      \\
                       & 300           & 0.25  &   1.33 & 1150  \\   
                       \midrule
\multirow{5}{*}{CaH$_6$}  & 150**\tnote{**}        & 0.58  &  2.97 & 964      \\ 
                       & 160           & 0.41  &  1.96 & 1204    \\ 
                       & 170           & 0.40  & 1.89 & 1419 \\ 
                       & 180           & 0.39  &  1.83 & 1249  \\ 
                       & 190           & 0.39  &  1.84 & 1333 \\ 
                         \bottomrule
\end{tabular}
\begin{tablenotes}
\item[*]*Extrapolation obtained by scaling the Eliashberg function at 129 GPa by a factor of 1.5. Extrapolated $T_c=353$ K.
\item[**]**Without accounting for the anharmonicity
\end{tablenotes}
\end{table}

 \begin{table}
\caption{Debye temperatures ($T_D$) for some superconducting metals, intermetallic compounds and compressed polyhydrides.}
\centering
\begin{tabular}{cccc}
\toprule
\textbf{Compound} & $\bm T_c$\textbf{, K} & \textbf{EPC parameter ($\bm\lam$)} & $\bm T_D$\textbf{, K}\\
\midrule
Hg~\cite{33} &  4.16 & 1.0 -- 1.6 & 72\\
Nb~\cite{33} & 9.22 & 0.82 -- 1.05 &  277\\
Nb$_3$Sn~\cite{46} &  17.9 & 1.6 -1.8 & 270\\ 
ThH$_{10}$ (170 GPa) &  161 & 1.65 & 1350\\
YH$_9$ (205 GPa)~\cite{12,kong_methods} &  235 & 2.66 & 1275\\
(La,Ce)H$_9$ (123 GPa)~\cite{chen3_methods,48} & 190 & 2.27 & 1107\\
\bottomrule
\end{tabular}
\label{table_D}
\end{table}

\begin{figure}[H]
\centerline{\includegraphics[width=0.4\columnwidth]{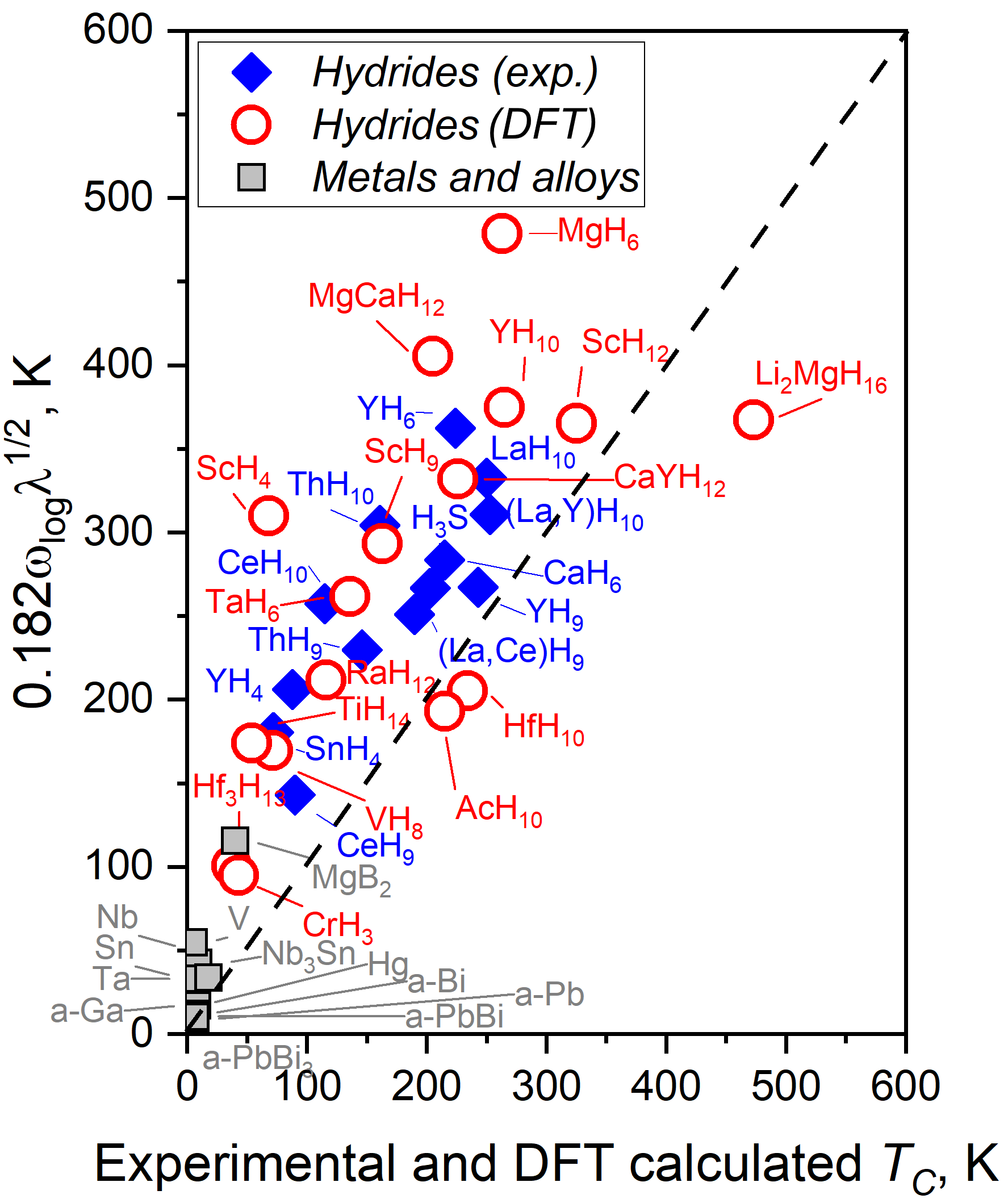}}
\caption{ Comparison of experimentally observed and DFT calculated critical temperatures $T_c$ for various superconductors with Allen-Dynes empirical strong coupling estimate  $T_c^{\mathrm{emp}}=0.182\omega_\mathrm{log}\sqrt{\lambda}$. Observe that for all experimentally studied hydrides $T_c < T_c^{\mathrm{emp}}$. }
\label{half_supp} 
\end{figure}

\end{document}